\documentclass[12pt]{article}

\pdfoutput=1
\usepackage{amssymb}
\usepackage{amsmath}
\usepackage{latexsym}
\usepackage[usenames]{color}
\usepackage{cite}
\usepackage{setspace}
\usepackage[margin=2.0cm]{geometry}
{}

\usepackage{hyperref}
\newcommand{\arXiv}[2]{\href{http://arxiv.org/abs/#1}{{\tt arXiv:#2}}}
\newcommand{\hep}[2]{\href{http://arxiv.org/abs/#1}{{\tt #2}}}

\begin{document}

\begin{center}
$$$$
{\Large\textbf{\mathversion{bold}
Semiclassical $SL(2)$ Strings on LLM Backgrounds
}\par}

\vspace{1.5cm}

\textrm{Minkyoo Kim \,\ \&\   \,\ Hendrik J.R. van Zyl}
\\ \vspace{1.5cm}
\footnotesize{\textit{
National Institute for Theoretical Physics,\\ School of Physics and Mandelstam Institute for Theoretical Physics, \\ University of the Witwatersrand, Wits 2050, \\ South Africa\\
}  
\vspace{8mm}
}
\textrm{E-mail: minkyoo.kim@wits.ac.za, hjrvanzyl@gmail.com}

\par\vspace{1.5cm}

\textbf{Abstract}\vspace{2mm}
\end{center}
We study semiclassical string solutions that live on white regions of the LLM plane for a generic LLM geometry. These string excitations are labelled by conserved charges $E, J$ and $S$ and are thus holographically dual to operators in the $SL(2)$ sector of $\mathcal{N} = 4$ super-Yang Mills made up of covariant derivatives acting on complex scalar fields $Z$.  On the other hand, the LLM geometry itself is dual to an operator consisting of ${\cal O}(N^2)$ $Z$-fields so that the operators dual to our solutions, containing both the stringy excitation and background, are non-planar.  In an appropriate short string limit we argue that the string solution we find should be dual to a localised $SL(2)$ excitation in the gauge theory language.  This allows us to perform a non-trivial check of the recent proposal that the dynamics of localised excitations should be identical, up to a rescaling of the 't Hooft coupling, to the dynamics of those same excitations in the $AdS_{5} \times S^{5}$ background.  

\setcounter{page}{1}
\newpage

\section{Introduction}

Semiclassical string solutions on $AdS_{5} \times S^{5}$ have provided invaluable data in the AdS/CFT correspondence \cite{AdSCFT, Witten, Gubser}.  The duality provides a one-to-one and onto map between gauge invariant operators of ${\cal N}=4$ super Yang-Mills theory and string theory states on an $AdS_5 \times S^5$ background.  The energy of a stringy excitation, for example, should match exactly with the operator dimension of the dual gauge invariant operator.

In the planar limit, we have several nontrivial matches between gauge theory and string theory. The single trace operators which are made of complex scalars could map to spin chain models \cite{Zarembo}, which are obtained from the dilatation spectrum for a specific sector, and are classified by Bethe eigenstates through number of flipped spins in an pseudo-vacuum state. The fundamental excitation is nothing but a single flipped spin state which is dual to a giant magnon \cite{GiantMagnons}, a rotating solitonic string on $S^{2}$. 
Another example is the GKP folded string \cite{GKP} in $AdS_{3} \times S^{1}$ which is holographically dual to the composite operator
\begin{equation}
Tr(D_{+}^{s_1} Z^{j_1} D_{+}^{s_2} Z^{j_2}  \cdots D_{+}^{s_n} Z^{j_n}).  \label{sl2AdSOps}
\end{equation} 
This operator forms part of the $SL(2)$ subsector of planar ${\cal N}=4$ SYM and is built from lightcone covariant derivatives ${\cal D}_{+}$ acting on complex scalars $Z$.

It is by now well established that the planar AdS/CFT is integrable \cite{IntRefs}.  This makes the semiclassical analysis particularly helpful since a key idea of integrability is the existence of the exact $S$-matrix.  Through the AdS/CFT correspondence the $S$-matrix should be read off from the string side as well as the gauge theory side.  The leading strong coupling piece of the $S$-matrix is computed as the phase shift for a classical string scattering process \cite{GiantMagnons}, and the worldsheet scattering can be directly computed from appropriate flat space limits of the string background \cite{wssc}. Furthermore, the leading finite size energy corrections of string states can be calculated from the string sigma model, and those can be compared to the L\"uscher's corrections \cite{Luscher, Luscher1} which can also be perturbatively obtained from the thermodynamic Bethe ansatz \cite{TBA1, TBA2, TBA3}. Moreover, classical string theory on $AdS_5 \times S^5$ itself can be reformulated through the classical spectral curve \cite{ClassSCurve}.  The generalisation of this, the quantum spectral curve \cite{QuantumSCurve}, contains all finite size effects so that the full spectrum may be extracted.

Beyond the planar limit the picture becomes more complicated.  In particular one may consider configurations that are massive enough that their backreation on the $AdS_5 \times S^5$ spacetime can no longer be ignored such as the bound states of many $D3$-branes that give rise to the $\frac{1}{2}$-BPS and regular LLM geometries \cite{LLM}.  In these setups there exist states that exhibit inelastic scattering and whose $S$-matrix does not satisfy the Yang-Baxter equation \cite{Anomalous, Inelastic, IntegrableSub}. 
String theory on these backgrounds thus cannot be integrable in general.  A recent proposal \cite{LLMMagnons} has claimed, however, that this does not imply that integrability is completely absent in these non-planar setups and all that is required is a procedure to restrict to the appropriate subsectors of the theory. 

Let us briefly review this proposal.  Its setting considers operators that have bare dimensions scaling as ${\cal O}(N^2)$.  Due to the large number of fields making up the operator it lies beyond the planar limit and many of the approximations associated with the planar limit are no longer valid \cite{Schur, Nonplanar2, Nonplanar3}.  In particular one can no longer focus solely on the planar diagrams.  An appropriate framework in which to study these problems is the Schur polynomial basis that makes heavy use of representation theory in order to sum all diagrams and not just the planar contributions \cite{SchurR1, SchurR2, SchurR3, SchurR4, SchurR5, SchurR6, SchurR7, SchurR8}.  This basis provides a very natural way to describe the operators dual to strings on LLM geometries.  Explicitly these are dual \cite{LLM, Reflecting, Anomalous, IntegrableSub, LLMMagnons, Geometries, Schur,StringAttached2, StringAttached3} to restricted Schur polynomials labelled by Young diagrams that organise the fields making up the dual gauge invariant operator.  These Young diagrams organise the ${\cal O}(N^2)$ fields dual to the bound state of $D3$-branes that backreact to form the background LLM geometry as well as the ${\cal O}(\sqrt{N})$ fields that are dual to string excitations on the background.  

A remarkable simplification occurs when the excitation fields are restricted to a single Young diagram corner isolated by at least ${\cal O}(N)$ boxes from the rest of the corners.  Under the action of the large $N$ dilatation operator these localised operators do not mix with background fields nor excitation fields located at other corners (these mixing terms are suppressed by a factor of $\frac{1}{N}$).  In fact, the claim of \cite{LLMMagnons} is that the action of the dilatation operator on these localised excitations is identical, up to a rescaling of the 't Hooft coupling,\footnote{Whenever a factor of $\lambda = g_{YM}^2 N$ appears it needs to be rescaled as $\lambda' = g_{YM}^2 N_{eff} = \lambda \frac{N_{eff}}{N}$ where $N_{eff}$ is the weight of the corner on the Young diagram where the excitations are attached (see also \cite{Miller1}, \cite{Miller2}, \cite{LinZeng} that reach similar conclusions.}
to its action on the same excitations in the trivial $AdS_{5} \times S^{5}$ background. Integrability should thus be a feature within these subsectors.

The checks of \cite{IntegrableSub} have already provided non-trivial evidence for the proposal of \cite{LLMMagnons} at both weak and strong coupling. The key point to mention is that, for these checks, there is the $su(2|2)^2$ residual symmetry which places significant restrictions on their spectrum and $S$-matrix.  This was first shown via an elegant argument in \cite{su22S} for the planar, elastic scattering case and later extended in \cite{Reflecting, Anomalous} to the inelastic scattering involving a magnon attached to a maximal and non-maximal giant giant graviton respectively.  
It is thus important that checks of the proposal are also performed outside of the $SU(2|2)^2$ sector.
 
It is for this reason that we would like to study the $SL(2)$ sector of nonplanar ${\cal N}=4$ SYM on an LLM geometry.  Our primary focus is on the string theory side of the correspondence, where we may compare with several works in the planar limit such as \cite{FrolovTseytlin, Basso, Basso2, Gromov, Gromov2, LargeWind} and see if the rescaled coupling proposal of \cite{LLMMagnons} is manifest in the spectrum of strings dual to localised operators.  
We hope furthermore to gain some insight into how the subsectors may be isolated on the string theory side.

The paper is organised as follows.  In section \ref{sec2} we review some relevant background focussing on the LLM geometry and the gauge theory prescription for localising operators.  In section \ref{sec3} we solve the equations of motion, discuss the boundary conditions of the states we are interested in and find expressions for the conserved charges in integral form.  We also briefly present some new insights into the use of the classical string length as a generating function for the classical conserved charges and discuss the region of parameter space where our string solutions may be trusted as good approximations to the strings dual to localised operators.  In section \ref{sec4} we investigate short strings in the white regions on the LLM plane and show how the classical result exactly match the proposal.  We also consider the long string limit of the solutions and conclude with some final remarks in section \ref{discussion}.

\section{Background \label{sec2}}

\underline{\textbf{LLM Geometries}}: The LLM metric \cite{LLM} is given by 
\begin{equation}
ds^2 = -y(e^G + e^{-G})(dt + V_i d x^i)^2 + \frac{1}{y(e^G + e^{-G})}(dy^2 + dx^i dx^i) + y e^G d\Omega_3 + y e^{-G} d\tilde{\Omega}_3.  
\end{equation}
The factors in the metric are all determined by a single function $z(x^1, x^2, y)$ as
\begin{equation}
z = \frac{1}{2}\tanh(G), \quad y \partial_y V_i = \epsilon_{ij} \partial_j z , \quad y (\partial_i V_j - \partial_j V_i) = \epsilon_{ij} \partial_y z
\end{equation}
while the function $z$ is sourced by a Laplace equation
\begin{equation}
\partial_i \partial_i z + y \partial_y \frac{\partial_y z}{y} = 0.  
\end{equation} 
Regularity requires that $z = \pm \frac{1}{2}$ on the LLM plane $(y=0)$ and thus the LLM geometries are specified completely by a coloring of the LLM plane with $z = \frac{1}{2}$ colored white and $z = -\frac{1}{2}$ colored black.

The geometries we are interested in have an LLM plane coloring of concentric rings of alternating color and thus we will find it useful to transform to radial coordinates $\left\{ x^1, x^2\right\} \rightarrow \left\{ r, \phi  \right\}$ so that $V_r = 0$ and $V_{\phi} = V_{\phi}(r)$.  For each such an LLM geometry there exists a direct mapping to a Schur polynomial labelled by a Young diagram, $B$ \cite{LLM, Geometries}.  The lengths of the sides of the Young diagram are mapped to the areas of the rings.  A vertical edge is mapped to a black ring while its adjacent horizontal edges are mapped to neighbouring white rings. 

Here are some useful identities for the concentric ring LLM geometries.  When we approach the LLM plane we find
\begin{eqnarray}
e^G & = & y e^{G_{-}(r)} + {\cal O}(y^3) = y \sqrt{\frac{\partial_r V_{\phi}}{2r}} + {\cal O}(y^3)  \ \ \ \textnormal{for} \ \ z = -\frac{1}{2}, \\
e^G & = & \frac{e^{G_{+}(r) }}{y} + {\cal O}(y) =  \frac{1}{y}\sqrt{-\frac{2r}{\partial_r V_{\phi}}}  + {\cal O}(y) \ \ \ \textnormal{for} \ \ z = \frac{1}{2}.  
\end{eqnarray}
The function $V_{\phi}$ may be written as
\begin{equation}
V_{\phi}(r) = \sum_{r_i < r}  c_i \frac{r_i^2}{r^2 - r_i^2} + \sum_{r_i > r}   c_i \frac{r^2}{r_i^2 - r^2} + {\cal O}(y^2)
\end{equation}
where $c_i = -1$ for the white rings ($z = \frac{1}{2}$) and $c_i = 1$ for the black rings ($z = -\frac{1}{2}$). \\ \\
\underline{\textbf{Localised Excitations}}:  In \cite{LLMMagnons} the so-called localised excitations of a background Schur polynomial $\chi_B(Z)$ were defined.  We illustrate the prescription here for excitations containing $Z$, $Y$ and $X$-fields.  Suppose the excitation is written in the Schur polynomial basis
\begin{equation}
O = \sum_{n} c_n \chi_{R_n, (r_n, s_n, t_n)}(Z, Y, X).  \label{PlanarOp}
\end{equation}
In this notation the $c_n$ are coefficients and the $r_n$, $s_n$ and $t_n$ are Young diagrams that organise the Z-fields, the Y-fields and the X-fields respectively.  The Young diagrams $R_n$ organise all the fields.  The background Schur polynomial consists of $Z$-fields and may have many possible corners to which boxes may be attached to yield a valid Young diagram.  If we wish to localise (\ref{PlanarOp}) at a given corner, say corner $0$, of the background $B$ this is given by
\begin{equation}
O_{B_0} = \sum_{n} c_n \chi_{ (R_{B_0})_n, ((r_{B_0})_n, s_n, t_n)}(Z, Y, X).  \label{LocalisedOp}
\end{equation}
It is vital that the corner $0$ of $B$ is a distant corner i.e. all other possible locations where one may add excitations are ${\cal O}(N)$ boxes away.  Under the action of the dilatation operator this property ensures that there is no mixing between the boxes making up the background and the excitation.  Note that there are only two changes - the diagram organising the $Z$-fields and the diagram organising all the fields need to be enlarged to accommodate the $Z$-fields from both the background and the excitation.  The coefficients and the other Young diagrams are not altered.  The generalisation to restricted Schur polynomials with additional restriction labels is obvious.  The $SL(2)$ operators we are interested in can also be described using the Schur polynomial basis, see \cite{sl2Gauge} for an example.  These can thus be localised at any corner of a background Schur polynomial by means of the same prescription. 

The one-loop computation of \cite{LLMMagnons} as well as the $SU(2|3)$ two-loop weak coupling and $SU(2)$ finite size strong coupling results \cite{IntegrableSub} indicate that the dynamics of localised excitations is related to the dynamics of the original excitation (with no background) in a remarkably simple way
\begin{equation}
D O_{B_0} = \left. \left( D O \right)_{B_0} \right|_{\lambda \rightarrow \lambda r_0^2} \label{DAction}
\end{equation}
where $r_0 = \sqrt{\frac{N_{eff}}{N}}$ and $N_{eff}$ is the weight of the box appearing at corner $0$ of the background diagram $B$.  The dynamics are thus identical up to a rescaling of the 't Hooft coupling.  It is this proposal that we wish to check in this work.

The dual string theory description of localised $SU(2)$ operators was already developed in \cite{IntegrableSub}.  The dispersion relation for the infinite size dyonic giant magnons matches perfectly with the strong coupling limit of the exact result \cite{Dyonic}.  The identification between strings and dual operators can be done by simply comparing the strong coupling exact anomalous dimensions to the string dispersion.

In the coming sections we will be interested in developing the dual string theory description for operators in the $SL(2)$ sector with an LLM background.  We are not aware of an existing exact dispersion relation for this sector so that the identification of dual operator is not as straightforward as for the $SU(2)$ sector.   We will argue, by specifying some criteria that a localised excitation should satisfy, that the solutions we find do not correspond to localised operators in general and can at best approximate them.  We are able to extract analytic results in two limits - long strings and short strings.  For the short strings we will show that we are able to satisfy all the criteria for a localised excitation and perform a non-trivial check of the proposal of \cite{LLMMagnons}.

\section{String Solutions \label{sec3}}

We will work with solutions restricted to the white ($z=\frac{1}{2}$) regions of the LLM plane.  We consider the Nambu-Goto action
\begin{equation}
S_{NG} = \frac{\sqrt{\lambda}}{2\pi} \int d\sigma d\tau \sqrt{ (\dot{X}\cdot X')^2 - \dot{X}^2 X'^2       }.
\end{equation}
The ten equations of motion can be solved by restricting to 
\begin{equation}
y = 0 \ \ ; \ \ \theta_2 \ , \ \theta_3 = {\rm constant} \nonumber
\end{equation}
and setting
\begin{equation}
t = \kappa \tau \ \ ; \ \ \phi = \alpha \tau \ \ ; \ \ \theta = \omega \tau .  \label{SL2Sols} 
\end{equation}
There is no restriction on $r(\tau, \sigma)$ coming from the equations of motion but it can be fixed by choosing a particular gauge.  We choose conformal gauge which implies 
\begin{equation}
(r'(\sigma))^2 = -(\kappa + \alpha)^2 r^2 + e^{2 G_{+}}(\kappa^2 - \omega^2 + 2\kappa(\kappa + \alpha)V_{\phi} + (\kappa + \alpha)^2 V_{\phi}^2). \label{rPrime}
\end{equation}
Aside from the condition for imposing conformal gauge there is no change compared to the solution of \cite{FrolovTseytlin} for the $AdS_5 \times S^5$ background.  For the $AdS_5 \times S^5$ setup these are dual to the $SL(2)$ gauge invariant operators (\ref{sl2AdSOps}).  We will return to the question of whether these solutions are dual to localised excitations for a generic LLM geometry shortly. \\ \\
\underline{\textbf{Boundary conditions}}: We need to mindful of the fact that (\ref{rPrime}) must remain positive and that the constants in (\ref{SL2Sols}) must remain real and finite.  This places restrictions on the allowed choices for the constants and the range for $r$.  It turns out that there are no points where $r'(\sigma) \rightarrow \infty$ so that our boundary conditions are the points where $r'(\sigma) = 0$.  Any radius that is a pole in $V_{\phi}$ satisfies this condition.  This can be seen by making the substitution
\begin{eqnarray}
V_{\phi}(r) & = & \frac{r_0^2}{r^2 - r_0^2}(1 + \bar{V}_{\phi}(r) ) \nonumber \\
\Rightarrow  \left. (r'(\sigma))^2 \right|_{r = r_0} & = & r_0^2 (\alpha + \kappa)^2 \bar{V}_{\phi}(r_0) = 0.  
\end{eqnarray}
This implies that our solutions are always confined to a single white region of the LLM plane.  Depending on our choice for the constants $\kappa, \alpha, \omega$ there may be another turning point for the solution.  For a generic LLM geometry it is not possible to determine the turning point, $r_m$, in terms of these constants.  We may, however, use the trick of \cite{IntegrableSub} to rather determine one of the constants in terms of the desired turning point.  We find
\begin{equation}
\omega^2 = (\kappa + (\alpha + \kappa) V_{\phi}(r_{m}) )^2 + \frac{1}{2}(\alpha + \kappa)^2 V_{\phi}'(r_{m}). \label{omegaSol}
\end{equation}
It will turn out that, for $\kappa > 0$, we need $\omega < 0$ to give a positive charge $S$.  There is an exception to this when we set $\alpha = -\kappa$ since then (\ref{omegaSol}) does not allow for an arbitrary endpoint to be chosen.  In this case we find
\begin{equation}
(r'(\sigma))^2 = -\frac{2 r (\kappa^2 - \omega^2)}{V_{\phi}'(r)},
\end{equation}
and the only sensible solutions are when the string is pinned to one of the rings i.e. to one of the geodesics on the LLM plane.  This can be understood since operators constructed from $Z$'s rotate along the geodesics at the speed of light \cite{GiantMagnons}.  By setting $\alpha = -\kappa$ we are stating that the dual operator is predominantly constructed from $Z$'s and must thus be pinned to a geodesic.  We still have the freedom to specify which geodesic so that these limit solutions can exist on any of the ring edges. 

Away from the $\alpha = -\kappa$ limit there is a turning point at a value $r_m > r_0$ in a white region of the LLM plane.  If we take $r > r_m$ we find that (\ref{rPrime}) becomes negative so that the string segment is confined to $r_0 \leq r \leq r_m$.  As we run over the range of $\sigma$ we may run over this interval in $r$ multiple times.  

One of the three constants parametrising the solution (\ref{SL2Sols}) needs to be chosen so that
\begin{equation}
2\pi = \int_{0}^{2\pi} d\sigma = 2n \int_{r_{0 }}^{r_{m}} \frac{1}{r'(\sigma)} dr \ \ \ ; \ \ \  n \in Z \label{PerRestr}
\end{equation}
We choose to fix $\kappa$ from the restriction (\ref{PerRestr}).  One could consider more general strings made up of stitching together string segments that are parametrised by different sets of constants i.e. different values for $E,J$ and $S$.  The dispersion relation we will find for the individual segments may be combined to yield the dispersion relations for these more general cases. 

We find the following explicit expressions for the conserved charges
\begin{eqnarray}
\frac{2 \pi}{\sqrt{\lambda}} E & = & \int d\sigma \frac{-(\kappa + \alpha) r^2 + e^{2 G_{+}}(1 + V_{\phi})(\kappa + (\kappa + \alpha)V_{\phi})   }{e^{G_{+}}\sqrt{ -(\kappa+\alpha)^2 r^2 + e^{G_{+}}(\kappa^2 - \omega^2 + 2\kappa(\kappa+\alpha) V_{\phi} + (\kappa+\alpha)^2 V_{\phi}^2 )     }} r'(\sigma), \nonumber \\ 
\frac{2 \pi}{\sqrt{\lambda}} J & = & \int d\sigma \frac{-(\kappa + \alpha) r^2 + e^{2 G_{+}}V_{\phi}(\kappa + (\kappa + \alpha)V_{\phi})   }{e^{G_{+}}\sqrt{ -(\kappa+\alpha)^2 r^2 + e^{G_{+}}(\kappa^2 - \omega^2 + 2\kappa(\kappa+\alpha) V_{\phi} + (\kappa+\alpha)^2 V_{\phi}^2 )     }} r'(\sigma), \label{CCharges} \\
\frac{2 \pi}{\sqrt{\lambda}} S & = & \int d\sigma \frac{-\omega e^{G_{+}}  }{\sqrt{ -(\kappa+\alpha)^2 r^2 + e^{G_{+}}(\kappa^2 - \omega^2 + 2\kappa(\kappa+\alpha) V_{\phi} + (\kappa+\alpha)^2 V_{\phi}^2 )     }} r'(\sigma).  \nonumber
\end{eqnarray}
In our computation we always change from an integral over $\sigma$ to an integral over $r$ as in (\ref{PerRestr}). 

A surprising feature of these conserved charges that we find rather puzzling is the fact that when we set $\alpha = 0$ we only find $J = 0$ when the geometry is $AdS_5 \times S^5$, despite the absence of an angular rotation by the string.  For us this signals that our solutions are incomplete and we thus need to be careful in interpreting our solutions - selecting appropriate boundary conditions may not guarantee that its dual is a localised operator of the form\footnote{For example, the dual operator may contain ${\cal O}(1)$ bosonic impurities with algebra central extensions \cite{su22S, GiantMagnons}.} (\ref{sl2AdSOps}).  \\
The most natural candidate for the complete solution, based on the mapping from the $r, y$ coordinates to $\rho$ (as used in \cite{FrolovTseytlin}), is to generalise the ansatz to $y = y(\sigma)$ so that the strings are allowed to lift off the LLM plane.  This would allow us to tune the values for the conserved charges without introducing additional ones.  This ansatz does make the computation much more involved and, since we are already able to test the proposal of \cite{LLMMagnons} in a non-trivial way, we postpone this general analysis to future work.  \\ \\
\underline{\textbf{Identitites}}: Before we proceed we note some interesting and useful identities.  The conserved charges (\ref{CCharges}) obey several restrictions that may be derived without evaluating any integrals.  The following differential equations are satisfied (already on the level of the integrand)
\begin{equation}
\kappa \partial_\mu E + \alpha \partial_\mu J + \omega \partial_\mu S = 0 \ \ ; \ \ \mu = \kappa, \alpha, \omega . \label{DERestr} 
\end{equation}
We also observe that
\begin{equation}
\kappa E + \alpha J + \omega S = 2 n \int_{r_{0}}^{r_{m}} L_{NG} \equiv L
\end{equation}
so that the classical string length, as a function of $\kappa$, $\alpha$ and $\omega$, serves as a generating function for the conserved charges
\begin{equation}
E = \partial_\kappa L \ \ ; \ \  S = \partial_\omega L \ \ ; \ \ J = \partial_\alpha L. 
\end{equation}
We find this role played by the classical string length quite remarkable and it hints at the fact that the theory may possibly be rewritten in a much more efficient way.  We would also like to point out that similar relations appear, for example, for the giant magnon solutions of \cite{IntegrableSub}.  As far as we can see there is no a priori reason to expect these simple relations and it must be one of the special properties of the geometries we are studying.

A final important relation we find is
\begin{equation}
\frac{1}{\kappa} E + \frac{1}{\alpha} J + \frac{1}{\omega} S = 2 n (\kappa + \alpha)^2 \int_{r_0}^{r_{m}} \frac{-r^2 + e^{2 G_{+}}V_{\phi}(1 + V_{\phi})   }{\alpha \kappa e^{G_{+}}\sqrt{-r^2(\kappa + \alpha)^2 + e^{2G_{+}}(\kappa^2 - \omega^2 + 2\kappa (\kappa + \alpha)V_{\phi} + (\kappa+\alpha)^2 V_{\phi}^2)}   }.  \label{TDual}
\end{equation}
For the $AdS_5 \times S^5$ case the right-hand side is identically zero, as also pointed out in \cite{FrolovTseytlin}.  This restricts the form of the conserved charges significantly.  Assuming the right-hand side of (\ref{TDual}) to be zero this would imply that
\begin{equation}
E = -\frac{\kappa}{\alpha} J - \frac{\kappa}{\omega} S
\end{equation}
and, after plugging these into the differential equations (\ref{DERestr}), we can fix all of the charges up to an overall function
\begin{eqnarray}
S & = &  \sqrt{\frac{(1 + \eta)\kappa^2 - \alpha^2 \eta}{\eta(\kappa^2 - \alpha^2)} } f_S(\kappa, \eta),  \nonumber \\
J & = & \sqrt{\frac{\alpha^2}{\kappa^2 - \alpha^2}} f_J(\kappa, \eta), \\ 
\partial_\eta f_J  & = & \sqrt{\eta} \partial_\eta f_S ,
\end{eqnarray}
where we have defined $\eta \equiv \frac{\omega^2 - \kappa^2}{\kappa^2 - \alpha^2} > 0$ borrowing the notation of \cite{FrolovTseytlin}.  We emphasise that this follows just from using the differential equations (\ref{DERestr}), which hold in general, and the additional constraint that (\ref{TDual}) is zero. 

Notice that in this case when we set $\alpha = 0$ we indeed recover $J = 0$ as we would expect for the appropriate solutions.  The quantity on the righthand side of (\ref{TDual}) can thus be thought of as one of the measures for how well our solutions can approximate the dual localised operators.  This may also be anticipated in another way.  The factor appearing in the numerator of (\ref{TDual}) is exactly the same factor that appears in the numerator of $J$ when $\alpha = 0$ (\ref{CCharges}).  

We note that (\ref{TDual}) is zero up to second order in $\alpha \approx -\kappa$ so that there is an indication that up to this order our solutions can approximate the dual localised operators well.  

\section{Analytic string limits \label{sec4}}

In order to make progress we need to approximate the integrals computing the conserved charges.  \\ \\
\underline{\textbf{Long String}}: We first consider the long string limit, where the string stretches from the outermost ring on the LLM plane to approximately the radial boundary.  We need to ensure that all the poles and zeros are captured correctly when we approach this limit.  We thus make the substitution
\begin{equation}
V_{\phi}(r) = \frac{r_0^2}{r^2 - r_0^2}(1 + \bar{V}_{\phi}(  r   )   )
\end{equation}
and the following change of coordinates
\begin{equation}
r = \frac{r_0}{z} \ \ ; \ \ r_m = \frac{r_0}{f} \ \ ; \ \ \alpha \equiv -\kappa\bar{\alpha} \nonumber \\
\end{equation}
so that the integrals we need to evaluate all run from $z = f$ up to $z = 1$. 
In \cite{IntegrableSub} a systematic expansion of these kinds of integrals that capture all the linear logarithmic divergences was developed.  Making use of this we find the following small $f$ expressions for the conserved charges
\begin{eqnarray}
\log(S) & = & - 2 \log(f) + {\cal O}(1),  \nonumber \\
J & = & -\frac{n \sqrt{\lambda}}{\pi} r_0 \frac{\bar{\alpha}}{\sqrt{1 - \bar{\alpha}^2}}\sqrt{1 + \bar{V}_{\phi}(\infty)} \log(f)  + {\cal O}(1), \\
E- S - J & = & -\frac{n \sqrt{\lambda}}{\pi} r_0 \frac{\sqrt{1 - \bar{\alpha}^2}}{1 + \bar{\alpha}}  \sqrt{1 + \bar{V}_{\phi}(\infty)} \log(f)  + {\cal O}(1).  \nonumber
\end{eqnarray}
Working up to ${\cal O}(1)$ contributions these may be combined to yield
\begin{equation}
E - S =  \sqrt{ J^2 + \frac{n^2 \lambda}{4 \pi^2} r_0^2 (1 + \bar{V}_{\phi}(\infty))\log^2(S)} + {\cal O}(1).
\end{equation}
At first glance there is a rescaling of the coupling as $\lambda \rightarrow r_0^2 (1 + \bar{V}_{\phi}(\infty)) \lambda$.  However, this is exactly the dispersion relation in the appendix of \cite{LargeWind} with no rescaling.  The way to see this is to realise that 
$r_0^2 (1 + \bar{V}_{\phi}(\infty)) $ sums the area of all the black regions on the LLM plane - this is a fixed quantity for all LLM geometries!  From our perspective this is strong evidence that our string solutions do not describe localised excitations in general.  The intuitive expectation is that the correct dual strings should at the very least be sensitive to the radius at which they are defined.  If they are sensitive to the other radii this would signal a breakdown of the proposal of \cite{LLMMagnons}.  However, the solution we find here is not sensitive to any of the radii and thus cannot be thought of as a string dual to a localised operator.   \\ \\
\underline{\textbf{Short string}}:  Fortunately the short string limit, where $r_{m} \approx r_0$, turns out to be easier to interpret.  We take
\begin{equation}
V_{\phi}(r) = \frac{r_0^2}{r^2 - r_0^2} + \tilde{V}_{\phi}(r)
\end{equation}
where $\tilde{V}_{\phi}$ is always ${\cal O}(1)$ in the short string approximation.  Considering (\ref{omegaSol}) we notice that if we were to make an expansion around $r_{m} \approx r_0$ then the leading order piece of $\omega$ goes to infinity unless
\begin{eqnarray}
\alpha & = & -\kappa + \frac{\kappa}{r_0}(r_{m}- r_0) \beta(r_m)  \nonumber \\
\Rightarrow \omega & \approx & - \kappa \sqrt{1 + \beta(r_0)} + o(r_m - r_0). \nonumber
\end{eqnarray}
In the above $\beta(r_m)$ is an ${\cal O}(1)$ and positive number.  The $r_m$ dependence is included since this is the most general thing one could do.  Our leading order computations will be unchanged whether it is treated as a function of $r_m$ or a constant.  We find it convenient to make the change of coordinates
\begin{equation}
r = r_0 (\sqrt{1 + \epsilon^2 z}) \ \ ; \ \ r_{m} = r_0 \sqrt{1 + \epsilon^2}. \nonumber 
\end{equation}
Expanding the integrand around $\epsilon \approx 0$ and then integrating over $z$ yields a systematic expansion in orders of $\epsilon$.  The leading order terms in $\epsilon^2$ are
\begin{eqnarray}
S & = & \frac{\Lambda}{2} \sqrt{1 + \frac{1}{\beta(r_0)}}\epsilon^2 , \nonumber \\
E & = & \Lambda \sqrt{\frac{1}{\beta(r_0))}} + \frac{ \Lambda ( (3 + 4 \tilde{V}_{\phi}(r_0)) \beta(r_0) + (1 + 6 \tilde{V}_{\phi}(r_0))\beta(r_0)^2 - 2 \beta'(r_0)  ) \epsilon^2 }{8 \beta(r_0)^{\frac{3}{2}}}, \label{Charges} \\
J & = &  \Lambda \sqrt{\frac{1}{\beta(r_0))}} - \frac{\Lambda ( (1 - 4 \tilde{V}_{\phi}(r_0)) \beta(r_0) + (3 - 6 \tilde{V}_{\phi}(r_0))\beta(r_0)^2 + 2 \beta'(r_0)  ) \epsilon^2 }{8 \beta(r_0)^{\frac{3}{2}}}, \nonumber
\end{eqnarray}
where we have defined $\Lambda \equiv \frac{n}{2} r_0 \sqrt{\lambda}$.  This quantity clearly scales depending on which ring the string segments end on. These expressions may be combined to yield the dispersion relation
\begin{equation}
E^2 = J^2 + 2 S \sqrt{J^2 + \Lambda^2   }  \label{stringDispersion}
\end{equation}
valid up to leading order in $S$.  Note that the dispersion relation is the same as that found in \cite{Basso}, up to a rescaling of $\lambda \rightarrow r_0^2 \lambda$ for all relevant LLM geometries.  This fits precisely with the proposal of \cite{LLMMagnons} and we emphasise that this match happens in a non-trivial way.  The individual conserved charges depend on the LLM background specifics (\ref{Charges}).  We have two constants that can be tuned to give different values for the conserved charges.  It is only when we rewrite these constants as functions of two of the conserved charges that all the details of the LLM background specifics cancel and we are left with the dispersion relation (\ref{stringDispersion}). 

When we expand up to second order in $\epsilon^2$ we find an apparent deviation from the rescaling proposal
\begin{equation}
E^2 - \left(J^2 + 2 S \sqrt{J + \Lambda^2} + S^2 \frac{J^2 + 3 \Lambda^2 }{2J^2 + 2\Lambda^2}\right) =  \frac{2 S^2 \Lambda^2 \tilde{V}_{\phi}(r_0)  }{J^2} \label{Dispersion}
\end{equation}
when compared to the detailed expressions in \cite{Gromov}.  As we explained above this is likely a consequence of the fact that these string solutions are only approximately dual to localised operators.  The quantity (\ref{TDual}) is non-zero at this order which fits well with our intuition that it is one of the ways to measure whether we are dealing with a local excitation or not.  At higher orders in the expansion we always find deviations that scale as $\frac{S^n}{J^{n+k}}$ with $k \geq 0$.  If we were to take a $J \rightarrow \infty$ limit all the deviations are thus suppressed.  We thus claim that our solutions can be trusted as dual to localised operators provided that $\frac{S^{2}}{J^2} \ll 1$.  Up to this order the results (\ref{stringDispersion}), (\ref{Dispersion}) clearly confirm the proposal of \cite{LLMMagnons}.  

\section{Discussion \label{discussion}}
The notion that there exist sectors in non-planar super Yang-Mills theory that have identical dynamics compared to the planar limit is an intriguing one.  Not only would this extend our understanding of the non-planar theory significantly but it would provide a new, novel setting in which to develop the framework of integrability.  A detailed proof of this proposal is likely to be a difficult task and as a first step it is essential to find non-trivial checks of this proposal. 

In this paper we found string solutions on an LLM background that are characterised by an angular momentum and twist.  These are thus dual to gauge invariant operators constructed from covariant derivatives acting on products of a single complex scalar field.  We argued that these string states are not necessarily dual to localised operators and that additional conditions must be satisfied.  In the short string limit we were able to satisfy these conditions and perform a non-trivial check of the proposal of \cite{LLMMagnons}.  This check should be read in conjunction with the existing results in the $SU(2)$ and $SU(2|3)$ sector \cite{LLMMagnons, IntegrableSub}.  Together these are starting to form mounting evidence that the proposal is indeed correct though many checks can and should still be performed.  \\ \\

\begin{centerline} 
{\bf Acknowledgements}
\end{centerline} 

MK is supported by the South African Research Chairs
Initiative of the Department of Science and Technology and National Research Foundation
as well as funds received from the National Institute for Theoretical Physics (NITheP).
HJR is supported by a Claude Leon Foundation postdoctoral fellowship.  The authors would like to thank Robert de Mello Koch for many valuable discussions and input.

\end{document}